\documentclass[10pt,
reprint,
twocolumn,
amsmath,
amssymb,
aps,
prb,
superscriptaddress,
longbibliography]{revtex4-2}

\usepackage{graphicx}
\usepackage[T1]{fontenc}
\usepackage{dcolumn}
\usepackage{bm}
\usepackage{fancyhdr}

\usepackage[colorlinks,allcolors=blue]{hyperref}
\usepackage{xcolor}
\usepackage[utf8]{inputenc}

\usepackage{import}
\usepackage{hyperref}
\usepackage{textcomp}
\usepackage{siunitx}
\usepackage{upgreek}
\usepackage{transparent}
\usepackage{multirow}

\DeclareSIUnit\angstrom{\text{Å}}

\begin{document}

\preprint{APS/123-QED}

\title{Vectorial Reconstruction of Magnetic Order Parameters using Electron Magnetic Linear Dichroism}


\author{Jan Hajduček}
\email{jan.hajducek@ceitec.vutbr.cz}
\affiliation{CEITEC BUT, Brno University of Technology, Purkyňova 123, 612 00 Brno, Czech Republic}

\author{Jáchym Štindl}%
\affiliation{Institute of Physical Engineering, Brno University of Technology, Technická 2, 616 69 Brno, Czech Republic}

\author{Ján Rusz}%
\affiliation{Department of Physics and Astronomy, Uppsala University, Box 516, 75120 Uppsala, Sweden}

\author{Vojtěch Uhlíř}
\email{vojtech.uhlir@ceitec.vutbr.cz}
\affiliation{CEITEC BUT, Brno University of Technology, Purkyňova 123, 612 00 Brno, Czech Republic}
\affiliation{Institute of Physical Engineering, Brno University of Technology, Technická 2, 616 69 Brno, Czech Republic}


\date{\today}

\begin{abstract}

Determining the orientation of compensated magnetic order with nanometer resolution remains a major challenge because conventional magnetic probes primarily detect net magnetization. Here, we establish electron magnetic linear dichroism as a~quantitative probe of the spin axis in transmission electron microscopy. Explicit inclusion of vectorial core-level exchange splitting into mixed-dynamic-form-factor simulations accounting for dynamical diffraction enables the calculation of the momentum- and energy-resolved dichroic response for an arbitrary Néel-vector or magnetization orientation. The magnetic linear contribution can be separated from nonmagnetic anisotropy and exhibits a characteristic angular dependence that enables reconstruction of the spin axis from a finite set of momentum-resolved electron-energy-loss spectra. Simulations for the antiferromagnetic and ferromagnetic phases of cubic FeRh show that this sensitivity persists from parallel illumination to atomically confined probes and over a broad range of crystal orientations and thicknesses. Electron magnetic linear dichroism therefore provides a route toward nanoscale vector imaging of compensated magnetic order in antiferromagnets and related magnetic materials.

\end{abstract}

\maketitle

\textit{Introduction:} Nanoscale reconstruction of magnetic order remains a~central challenge in spintronics and quantum materials, particularly for compensated magnetic phases such as antiferromagnetic (AF) \cite{Jungwirth2016, Baltz2018} and recently described altermagnetic \cite{Smejkal2022} order. In these systems, the antiparallel alignment of neighboring atomic magnetic moments yields a~vanishing net magnetization, making them inaccessible to conventional magnetization-sensitive probes \cite{Cheong2020}. X-ray magnetic dichroism has become the workhorse for element-specific magnetic imaging, providing direct access to local spin and orbital moments through circular dichroism (XMCD) \cite{Carra1993} and to the N\'eel vector orientation through linear dichroism (XMLD)~\cite{VanderLaan1986}. XMCD probes the magnetization component along the X-ray helicity axis and enables imaging of ferromagnetic (FM) domains and textures such as skyrmions and domain walls~\cite{Nolting2010,Cascales-Sandoval2024DeterminationXMCD-PEEM,Legut2014InfluenceExperiments}. In contrast, XMLD arises from anisotropic transitions governed by the local exchange field and provides sensitivity to AF domains and spin axes~\cite{Arai2012,Schmitt2023,Luo2019,Elnaggar2020ProbingXMLD}. Recently, their combination has enabled vectorial reconstruction of altermagnetic order~\cite{Amin2024NanoscaleMnTe}, supported by \textit{ab initio} simulations of X-ray dichroism~\cite{Hariki2024Determination/math, Hariki2024X-Ray-MnTe}.

A complementary route to magnetic dichroism is provided by inelastic electron scattering in a~transmission electron microscope (TEM), commonly used for atomic-scale materials characterization. Momentum-resolved electron energy-loss spectroscopy (EELS) enables detection of electron dichroism: energy loss resolves electronic transitions, while momentum transfer encodes the chirality required for circular dichroism \cite{Hebert2003,Schattschneider2005} and the directional sensitivity underlying natural \cite{Gloter2000,Hebert2003, Yamaguchi2018} and magnetic linear dichroism \cite{Yuan1997}. This formal analogy between polarized x-ray absorption and momentum-resolved EELS led to the development of electron magnetic circular dichroism (EMCD) as the electron counterpart of XMCD~\cite{Schattschneider2006}. EMCD enables the detection of magnetic order with atomic resolution~\cite{Wang2018AtomicMicroscopy,Rusz2016MagneticResolution,Rusz2014prl,Ali2020,Idrobo2016,Ali2025VisualizingMicroscope}, including recently demonstrated sensitivity to ferrimagnetic \cite{Wang2018AtomicMicroscopy} and AF order \cite{Song2026} using atomically sized convergent electron probes. Despite these advances, the vectorial sensitivity of EMCD remains only partially explored. To date, experiments have mainly accessed the projection of the uncompensated magnetization, with simultaneous sensitivity to both in-plane and out-of-plane components demonstrated only in limited cases \cite{Song2017a}, and full vector reconstruction remains largely conceptual \cite{Song2021}.

While the circular form of electron dichroism is now well established for high-resolution magnetic characterization, the linear form of electron dichroism (ELD), sensitive to anisotropic transitions, has so far been explored mainly in nonmagnetic systems for directions transversal to the electron beam~\cite{Leapman1983,Browning1991,Botton2005,Sun2005,Radtke2006,Menon1999,Saitoh2006,Arenal2007High-angular-resolutionNitride,Yamaguchi2018, Guzman2026}, also known as natural ELD (ENLD). The magnetic extension of ELD, electron magnetic linear dichroism (EMLD), was only demonstrated in AF hematite~\cite{Yuan1997,VanAken2003}, providing a~spectroscopic signature of the Morin transition~\cite{Morin1950MagneticTitanium}, revealing the EMLD sensitivity to the N\'eel vector component along the electron beam. However, the development of EMLD as a~robust imaging contrast mechanism has been limited by detector sensitivity and the lack of a~predictive theoretical framework. Although advances in direct electron detection technology \cite{McMullan2014ComparisonMicroscopy} have largely resolved the former, the latter remains a~key challenge. 

\begin{figure*}[t]
\includegraphics[width = 2\columnwidth]{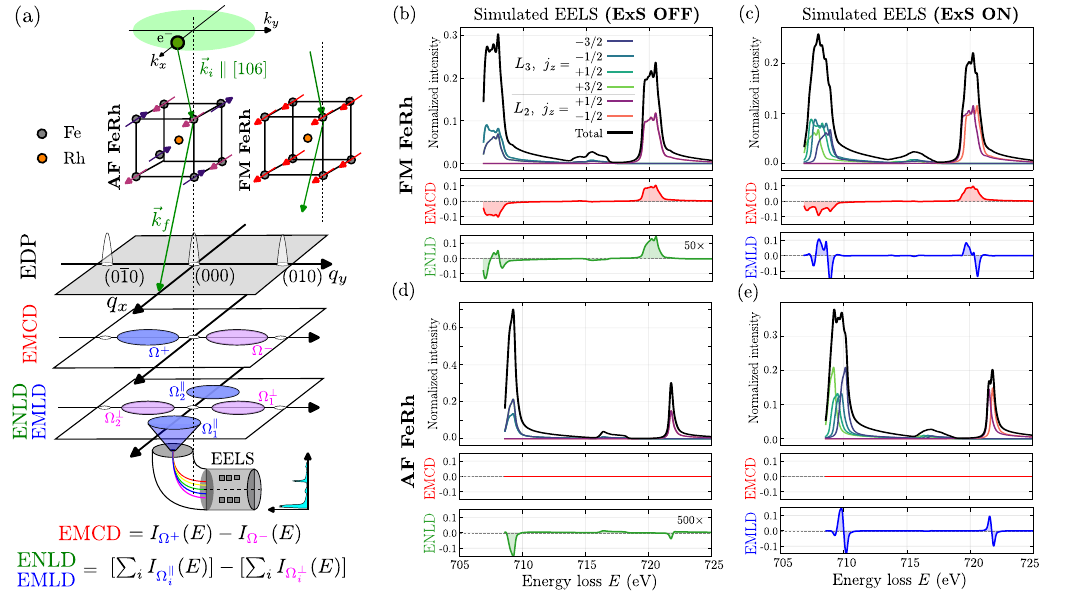}
\caption{\label{fig1:scat-geom} \textbf{Electron Magnetic Dichroism in TEM.} (a) Schematic of the EMD scattering geometry with incident electron beam along $\vec{k}_{i}$, slightly tilted from ZA direction, undergoing elastic and inelastic scattering in FeRh. Elastic scattering forms an~electron diffraction pattern (EDP) with indicated Bragg spots. Energy-filtered intensities $I_{\Omega}(E)$ are collected at symmetry-related aperture positions $\Omega$. Spectral differences are obtained by subtracting signals from $\Omega^{\parallel}$ and $\Omega^{\perp}$ for EMLD/ENLD, and from $\Omega^{+}$ and $\Omega^{-}$ for EMCD, which isolates the corresponding dichroic components. Simulated EELS spectra at the Fe $L_{3,2}$ edges, with color-coded $j_z$-resolved transitions and the integrated spectrum shown in black. Panels (b,c) correspond to the FM phase and (d,e) to the AF phase of FeRh. (c,e) include core-level exchange splitting (ExS) with apparent shifts in the $j_z$ levels, while (b,d) do not. The resulting spectral differences separate contributions from EMCD, EMLD, and ENLD. EMLD signal appears in both FM and AF phases when exchange splitting is present, while EMCD persists only in the FM case. 
}
\end{figure*}

In this Letter,  we establish the missing vectorial framework for EMLD. We show that the electron magnetic dichroism can be decomposed into its circular and linear terms and establish a~general route to vectorial reconstruction of magnetic order parameters. Within a~unified vectorial simulation framework, the distinct components of the dichroic signal are isolated and linked to the underlying magnetic degrees of freedom. Focusing on EMLD, we demonstrate that a finite set of momentum-resolved spectra is sufficient to reconstruct the spin axis in both compensated and uncompensated magnets. We illustrate the framework using FeRh, a prototypical metamagnet that undergoes a first-order transition from a low-temperature antiferromagnetic phase to a ferromagnetic phase above $\sim360$ K \cite{Maat2005, Lewis2016}. It has been extensively studied in thin film geometry using X-ray-based \cite{Baldasseroni2012, Baldasseroni2015} or TEM-based imaging techniques \cite{Almeida2020,Hajducek2026DislPRB, Hajducek2026EMCDPRM}. Its structurally similar but magnetically distinct phases provide a controlled setting for comparison of the linear and circular dichroic responses of compensated and uncompensated order. We further show that the EMLD response remains robust over broad ranges of beam orientation, sample thickness, and probe size. This promotes EMLD from a spectroscopic signature of magnetic anisotropy to a quantitative vector-sensitive probe of magnetic order.

\textit{Origin of EMCD and EMLD:} Fundamentally, EMCD and EMLD signals encode different properties of the magnetic order. EMCD is odd under reversal of the magnetic moment and therefore probes an oriented magnetization component. As EMLD is even under reversal, it measures the orientation of the magnetic axis rather than its sign, which makes it generally sensitive to both compensated and uncompensated magnetic order.

Experimentally, the scattering geometry underlying EMD is shown in Fig.~\ref{fig1:scat-geom}(a). An incident electron beam undergoes elastic and inelastic scattering in the FeRh crystal, forming a~diffraction pattern in the TEM back focal plane, with inelastic intensity localized around Bragg reflections. By selecting energy and momentum transfer in this plane, specific electronic transitions are probed. In transition metals, these correspond to core-level excitations (e.g., $2p \rightarrow 3d$ at the $L_{2,3}$ edges), where the unoccupied $3d$ states carry magnetic information. Spin–orbit coupling and core-level exchange splitting lead to anisotropic transition probabilities, projecting as EMLD, while spin-dependent valence-band asymmetry projects as EMCD.

We consider a~geometry with magnetic moments perpendicular to the incident beam (in-plane). EMCD is accessed via conjugate aperture positions $\Omega^{+}$ and $\Omega^{-}$ in the diffraction plane, typically under two- or three-beam conditions that enhance sensitivity to chiral transitions~\cite{Song2017a,Song2021}. In contrast, the linear dichroic component (EMLD/ENLD) is obtained by comparing intensities along orthogonal momentum-transfer directions, $\Omega^{\parallel}$ and $\Omega^{\perp}$, analogously to XMLD. Aperture positions are guided by calculated dichroic signal maps, while their size is maximized to ensure sufficient count rates without compromising momentum selectivity.

Examples of simulated EELS spectra, including individual $j_z$-resolved spectral components contributing to $L_2$ and $L_3$ edges, can be found in Fig.~\ref{fig1:scat-geom}(b-e) with and without core-level exchange splitting and for AF and FM phases of FeRh. The presence of spin polarization near the Fermi level, characteristic of FM ordering, directly leads to EMCD in EELS spectra as long as the beam splitter configuration conditions are met \cite{Hebert2003,Schattschneider2006}. Because valence-state polarization is present in the FM phase but absent in the AF phase, EMCD is detectable only in the FM phase, provided the electron interaction volume extends beyond individual atomic columns \cite{Song2026}.

Structural anisotropy, in turn, manifests in EELS as ENLD \cite{Hebert2003}. In cubic metallic magnets with weak structural anisotropy, such as FeRh, the ENLD response is small in both magnetic phases, unlike systems with strong orbital anisotropy, including transition-metal oxides, AF insulators, layered magnets, and compounds with large magnetocrystalline anisotropy \cite{Gong2017,Meer2023}. By including the local exchange field splitting, the degeneracy of the spin-orbit-split core sublevels is lifted, giving rise to the EMLD signal. As exchange splitting is present in both AF and FM phases, EMLD can be detected in both, as is the case for XMLD \cite{Stohr2006,Kuch2007}. Inclusion of exchange splitting is formalized in END MATTER. Although EMLD and ENLD are detected within the same aperture geometry, their magnitudes differ significantly, and they can be distinguished through their different dependence on magnetic order. Pure ENLD can be isolated by suppressing magnetic order, e.g., heating above the ordering temperature where EMLD vanishes. The structural ENLD contribution also remains unchanged when the spin axis is rotated, whereas EMLD follows the orientation of the Néel vector or magnetization. EMLD can therefore be isolated by comparing two states with different magnetic-axis orientations but identical crystal geometry, or by subtracting a reference spectrum acquired above the magnetic ordering temperature. This separation is fully analogous to XMLD-based protocols for disentangling magnetic and structural anisotropy \cite{Guo1994, Stohr2006}. Quantitative separation of EMCD, EMLD and ENLD contributions is evaluated in Supp. Note S1 \cite{EMLD-th-suppl}.

 \begin{figure}[b]
\includegraphics[width = 0.9\columnwidth]{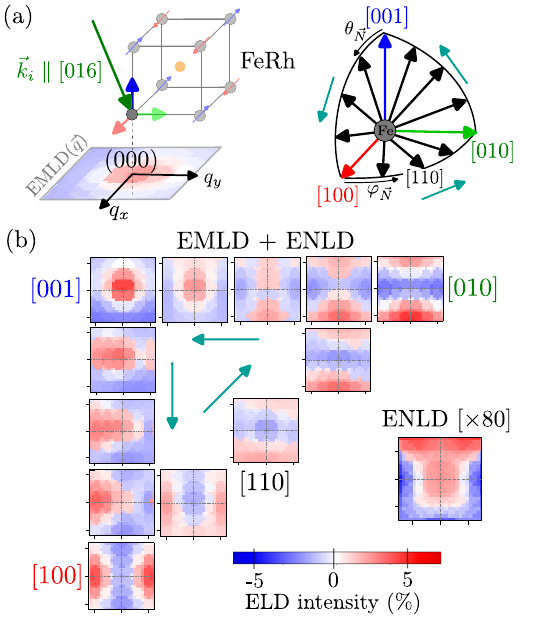}
\caption{\label{fig2:vect-dep}\textbf{Linear dichroic signal dependence on AF spin axis.} (a)~Schematic of the scattering geometry with indicated Néel vector orientations. (b) Integrated ELD intensity maps in the diffraction plane normalized by the original spectral intensity for a~given Néel vector orientation. The signal contains both EMLD and ENLD contributions, where the latter acts as a~spin-independent background. The signal is evaluated at each $\vec{q}$ position at the Fe $L_{3}$ edge in the momentum range of $(-0.6, +0.6)\text{G}$ from the (000) Bragg spot.}
\end{figure}

\begin{figure*}[t]
\includegraphics[width = 2\columnwidth]{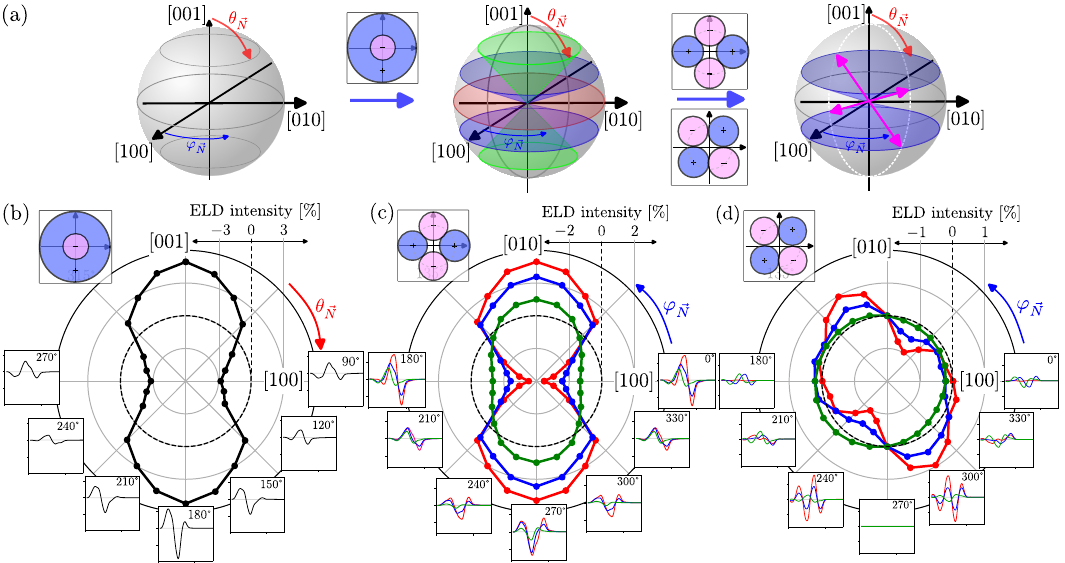}
\caption{\label{fig3:vect-recon}\textbf{EMLD-based vectorial reconstruction of the AF spin axis.} (a) Reconstruction scheme for an~unknown Néel-vector orientation. EMLD spectra acquired at selected aperture positions progressively constrain the possible spin-axis directions on the orientation sphere, yielding two symmetry-equivalent Néel-vector solutions (indicated by pink arrows) in the most general case. 
(b-d) Simulated EMLD spectral profiles (insets) and corresponding integrated dichroic intensity as a~function of $\theta_{\vec{N}}$ and $\varphi_{\vec{N}}$ for representative aperture configurations. The plots represent the full angular dependence of all possible EMLD responses obtained by subtracting the signal from the conjugate apertures indicated for each geometry.}
\end{figure*}


\textit{Vectorial EMLD simulations:} The inclusion of the core-level exchange splitting effectively introduces Zeeman-like energy shifts tied to the magnetic axis and as a~result, the EMLD signal becomes explicitly sensitive to the orientation of the magnetic moment. Correct treatment of the directional dependence is implemented into the simulation procedure from Refs.~\cite{Schattschneider2006, Rusz2007First-principlesFerromagnets} to form a~vectorial simulation framework, extending the mixed-dynamic-form-factor (MDFF) formalism to arbitrary magnetic-axis orientation. First-principles transition matrix elements are transformed by consistently rotating both the spin populations and the core-state angular-momentum basis. The resulting MDFF is then propagated through the full dynamical-diffraction calculation. This yields the energy- and momentum-resolved scattering cross section for any spin-axis orientation (for details see Supp.\ Note S2 \cite{EMLD-th-suppl} and END MATTER). 

This approach enables the calculation of EMCD and EMLD signal distributions in the diffraction plane, as well as their corresponding spectral profiles at each $\vec{q}$ point, for an~arbitrary spin-axis orientation. To illustrate this capability, we evaluate the normalized integrated signal distribution in the diffraction plane, defined as $\Delta I=\pm\int_{L_3}\vert\Delta I(E)\vert dE/\int_{L_2+L_3}I(E)dE$, with the sign chosen by the slope at the onset of the $L_3$ edge. 

The intensity maps corresponding to continuous rotation of the Néel vector $\vec{N}$ through the principal crystallographic directions of AF FeRh are shown in Fig.~\ref{fig2:vect-dep}. Due to the absence of net spin polarization, the response contains no EMCD and consists only of magnetic EMLD superposed on the much weaker structural ENLD background. The characteristic redistribution of positive and negative dichroic intensity provides a momentum-space fingerprint of the spin-axis orientation (Fig.~\ref{fig2:vect-dep}(b)). The spin axis follows the path $[100]\rightarrow[010]\rightarrow[001]\rightarrow[100]$, with a~smooth evolution of the signal between these orientations.

Different components of the Néel vector or magnetization can then be selectively probed using tailored aperture geometries~\footnote{To evaluate spectral profile for aperture defined by a~collection range $\Omega$, we calculate: $I_{\Omega}(E)=\sum_{\vec{q}_{i}\in \Omega} I_{\vec{q}_{i}}(E)$. To account for energy broadening, we evaluate: $I_{\Omega, \Delta E}(E)=Ltz_{\Delta E}(E)*I_{\Omega}(E)$, where $Ltz_{\Delta E}(E)$ is Lorentzian profile with the FWHM of $\Delta E$. Finally, the spectrum in $\Omega$ is normalized as: $I_{\Omega,\Delta E}^{\text{norm}}(E)=I_{\Omega,\Delta E}(E)/[\int_{E_{L_3 + L_2}}I_{\Omega,\Delta E}(E)dE]$, allowing the subtraction of spectral profiles and for arbitrary magnetic axis orientation.}. As demonstrated in Refs.~\cite{Yuan1997,VanAken2003}, sensitivity to the out-of-plane (OOP) component of the magnetic axis in EMLD is obtained by choosing the relative contributions of momentum-transfer components parallel and perpendicular to the magnetic axis, which is achieved by varying radii of concentric apertures. Complementarily, sensitivity to in-plane (IP) components is obtained by employing off-center aperture positions in the diffraction plane, enabling discrimination between orthogonal IP directions.

\textit{Dichroism-based vectorial reconstruction:} The evaluation of dichroic signal components by varying collection apertures enables the magnetic axis to be reconstructed from a finite set of spectra. The procedure is illustrated in Fig.~\ref{fig3:vect-recon}(a). For a known crystal structure, an OOP-sensitive aperture configuration is first used to determine the polar angle~$\theta_{\vec{N}}$. Subsequently, the IP orientation $\varphi_{\vec{N}}$ is obtained by comparing IP-sensitive EMLD spectra with their angular dependence simulated for the specific $\theta_{\vec{N}}$. In total, measurements at ten aperture positions enable full reconstruction of the spin-axis orientation. One aperture set determines the out-of-plane component, whereas two complementary in-plane-sensitive sets are required to uniquely determine the azimuthal orientation.

\begin{figure}[t]
\includegraphics[width = 0.92\columnwidth]{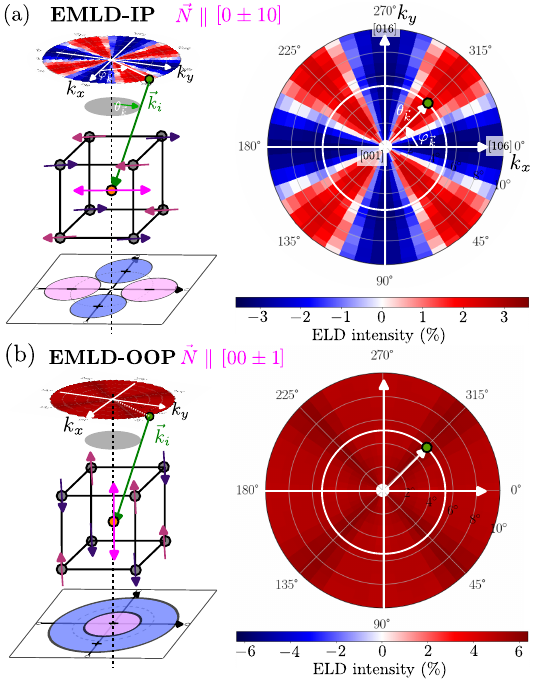}
\caption{\label{fig4:beam-dir-dep} \textbf{Crystal-orientation dependence of the EMLD signal in FeRh.} 
Schematic experimental geometries used to probe the (a) in-plane and (b) out-of-plane components of EMLD. Calculated EMLD signal maps as a~function of beam tilt magnitude and azimuth with respect to the [001] zone axis. Color scale encodes the sign and magnitude of the EMLD signal for each incident-beam direction and corresponding aperture configuration. Qualitatively similar beam-tilt dependencies of EMLD can be obtained for the FM case.}
\end{figure}

The angular dependence of the EMLD spectral profiles and integrated dichroic signal exhibits a~smooth harmonic response with extrema corresponding to orthogonal magnetization directions, as shown in Fig.~\ref{fig3:vect-recon}(b–d). The calculated profiles capture the full range of possible EMLD responses. In practice, the angular response need not be matched to an absolute calculated intensity, as the calibration against two experimentally accessible orthogonal reference states is sufficient to determine the relative position between the extrema. While the OOP-sensitive configuration is robust against IP rotation, the IP angular signal variation decreases with increasing OOP component, leading to reduced IP sensitivity in predominantly OOP geometries. 
The complete procedure in the most general case, without imposing other constraints on magnetic configuration, gives two equivalent solutions for the Néel vector (pink arrows in Fig.~\ref{fig3:vect-recon}(a)), which can be distinguished by the response of the corresponding OOP dichroic signal to sample tilting~\footnote{In practice, quantitative reconstruction can be performed relative to experimentally accessible reference extrema, typically obtained from orthogonal measurement geometries or saturated reference states, avoiding the need for absolute spectral-profile matching. While both spectral shape and corresponding integrated intensity contain angular information, the latter is expected to provide a~more experimentally robust observable.}.

\vspace{0.5cm}
\textit{Dichroic signal robustness}: A key advantage of using EMLD for vectorial reconstruction is its robustness across scales and parameter space. First, EMLD persists over a~broad range of crystal and electron beam orientations. Unlike EMCD, which requires near-ideal two- or three-beam conditions \cite{Schattschneider2006} (see Supp.\ Note~S3~\cite{EMLD-th-suppl}), the EMLD signal remains present across an extended angular range of incident directions $\vec{k}_i$ (see Fig.~\ref{fig4:beam-dir-dep}). Although its magnitude and sign vary with crystal orientation (Fig.~\ref{fig4:beam-dir-dep}(a) for IP component), the presented simulation framework allows its full predictive control. Notably, a small tilt away from the exact zone axis avoids the signal instability caused by dynamical diffraction, which justifies the use of slight off-normal beam incidence throughout this work. In contrast, the OOP component remains stable over the same angular range (Fig.~\ref{fig4:beam-dir-dep}(b)). Second, the EMLD signal is robust with respect to sample thickness, retaining both its sign and spectral shape (see Supp. Note~S4~\cite{EMLD-th-suppl}). Third, EMLD survives probe confinement from parallel beam down to atomically sized probes (for details on convergent beam EMLD simulations see Supp. Note S5 \cite{EMLD-th-suppl}). In contrast, EMCD in a compensated AF becomes detectable only with atomically confined probes capable of spatially resolving inequivalent sublattice contributions, thus placing stringent demands on experimental detection of AF order using EMCD.

Additionally, EMLD reproduces the characteristic harmonic dependence known from XMLD \cite{Arenholz2006, Stohr2006} and previous EMLD experiments~\cite{Yuan1997,VanAken2003}. The agreement between our spectral simulations and the dielectric-tensor formalism (Supp. Note S5 \cite{EMLD-th-suppl}) establishes a~direct conceptual connection between electron- and photon-based linear dichroism. This correspondence enables the transfer of established XMLD methodologies to the TEM, including quantitative determination of local magnetic-axis orientation and magnetocrystalline anisotropy. More generally, the complementary even and odd magnetic responses provided by EMLD and EMCD offer a possible route to distinguishing spin-axis anisotropy from momentum-dependent spin polarization in altermagnets, although material-specific calculations will be required. The details of the presented spectral simulations can be found in Supp. Note S6 \cite{EMLD-th-suppl}.

\textit{Conclusion}: We have established EMLD as a quantitative vector-sensitive probe of magnetic order in the TEM. Its linear-dichroic contrast survives AF compensation, can be separated from structural anisotropy, and enables reconstruction of the magnetic axis from a finite set of collection geometries. For optimized configurations, the calculated contrast reaches several percent, while the assumed 0.2 eV broadening is compatible with modern monochromated EELS. Together with its robustness against alignment, thickness, and probe confinement, this places experimental realization within reach of advanced aberration-corrected TEMs with direct electron detection. The method opens a route to nanoscale mapping of magnetic-axis orientation and anisotropy beyond the spatial reach of photon-based linear dichroism.

\vspace{0.5cm}
\textit{Acknowledgement:} This work was supported by the project TERAFIT No. CZ.02.01.01/00/22$\_$008/0004594. We acknowledge CzechNanoLab Research Infrastructure (ID 90251), supported by MEYS CR. J.R. acknowledges the support of the Swedish Research Council (grant no.\ 2025-04514) and the Knut and Alice Wallenberg Foundation (grant no.\ 2022.0079). The simulations were enabled by resources provided by the National Academic Infrastructure for Supercomputing in Sweden (NAISS) at the NSC Centre, partially funded by the Swedish Research Council through grant agreement no.\ 2022-06725.


\vspace{0.5cm}
All data and code used to generate the presented figures are available in the Zenodo repository \cite{ZENODO_EMLD}.

\bibliography{references-main-MAN}

\onecolumngrid 
\section*{End matter}
\twocolumngrid 

\appendix
\setcounter{equation}{0}
\renewcommand{\theequation}{A\arabic{equation}}

\textit{Implementation of vectorial EMLD}: This section summarizes the computational implementation used to evaluate EMLD for arbitrary orientations of the Néel (or magnetization) vector within the MDFF formalism. A consistent transformation of both the spin population and the core-level $j,j_z$ basis is required to capture the vectorial nature of the exchange-induced anisotropy.

Spin states from DFT~\cite{Blaha2020WIEN2k:Solids} are typically obtained for magnetization $\parallel [001]$ and are represented as $\vert ss'\rangle$ states separately for each sublattice in the AF case or combined for the FM case. Rotation of the spin space is accomplished through the spin-$\tfrac{1}{2}$ Wigner $D$-matrix, $D^{(1/2)}(\alpha,\beta,\gamma)$. In the absence of valence spin–orbit coupling (SOC), the spin population can be expressed as a~diagonal operator $\hat{M}=\text{diag}(M_{\uparrow \uparrow},M_{\downarrow \downarrow})$, which allows a~simple rotation:

\begin{equation}
    \hat{M}'_{ss'} = D^{(1/2)}_{ss'}(\alpha,\beta,\gamma)\, \hat{M}_{ss'} \, D^{(1/2)\dagger}_{ss'}(\alpha,\beta,\gamma).
\end{equation}

To ensure consistency between the rotated spin populations and the angular-momentum quantization axis, we also rotate the $\vert jj_z \rangle$ basis into $\vert jj_z \rangle_{\hat{z}\parallel\vec{M}}$. This transformation is achieved by expressing the core-to-valence transition elements from real-space basis states $\vert \vec{r}s\rangle$ into total-angular-momentum space through Clebsch–Gordan coefficients $C_{l m \tfrac{1}{2} s}^{j j_z'}$ and a~Wigner $D$-matrix $D^{(j)}{j_z' j_z}$:

\begin{align}
\langle \vec{r}s| j j_z \rangle_{\hat{z} \parallel\vec{M}}
&= \sum_{j_z'} \langle \vec{r}s | j j_z' \rangle \langle j j_z' | j j_z \rangle \notag \\
&= \sum_{j_z'} C_{l m \tfrac{1}{2} s}^{j j_z'} \, R_{js}(\vec{r}) \, Y_{lm} (\hat{r})\, 
D^{(j)}_{j_z' j_z}(\alpha,\beta,\gamma).
\label{eq:wigner_rot}
\end{align}

Here, the Euler angles $(\alpha,\beta,\gamma)$ align the local $j$-quantization axis with the Néel vector $\vec{N}$. The rotated quantization axis enables a~consistent application of exchange-split energy offsets for each $j,j_z$ component. In the MDFF evaluation, this transformation replaces the unrotated MDFF term $S(\vec{q},\vec{q}',E)$ as defined in \cite{Schattschneider2006, Rusz2007First-principlesFerromagnets} by its rotated counterpart as:

\begin{equation}
    S(\vec{q},\vec{q}',E) \mapsto S_{j,j_z}^{(\alpha,\beta,\gamma)}(\vec{q},\vec{q}',E)
\end{equation}

Exchange splitting and magnetic anisotropy are key to the origin of linear dichroism \cite{Mertins2001ObservationEffect, Ebert1996, Kunes2002X-raySplitting,Kunes2003}.
In XMCD, the local exchange field determines the spin-selective core excitations, while in XMLD its anisotropic component governs the angular dependence of the dichroic contrast~\cite{Kunes2004, Dhesi2002, VanDerLaan2011, Tesch2014, Arenholz2006, Arenholz2007}.

To capture these effects in the electron-scattering formalism, we impose the local spin quantization axis along the direction of magnetic order parameter and model the exchange-polarized $2p$ core states through projection-dependent energy shifts as: $\Delta E_{j,j_z} = \lambda_j j_z$, where $j_z$ is the projection onto the local quantization axis and $\lambda_j$ is an~effective exchange parameter for a~given $j$ (in practice obtained from either constrained DFT, atomic multiplet fits, or treated as a~tunable parameter).
For an~AF the two opposite FM sublattices cause sign reversal of the associated energy terms of the local spin polarization \emph{e.g.} $\Delta E_{j,j_z}^{(+)}=+\lambda_j j_z$ and $\Delta E_{j,j_z}^{(-)}=-\lambda_j j_z$, ensuring the correct sign reversal of the local spin polarization in the total MDFF sum. Operationally, these shifts are introduced by modifying the MDFF energy argument as:

\begin{equation}
S^{(\alpha,\beta,\gamma)}_{j,j_z}(\vec q,\vec q',E)\;\mapsto\; S^{(\alpha,\beta,\gamma)}_{j,j_z}(\vec q,\vec q',\,E-\Delta E_{j,j_z}).
\end{equation}

Combining the elements above, the simulated double-differential cross section for a~given orientation $(\alpha,\beta,\gamma)$ becomes

\begin{align}
\frac{\partial^2\sigma^{(\alpha,\beta,\gamma)}}{\partial\Omega\,\partial E}
&=\frac{4\gamma^2}{a_0^2}\frac{k_f}{k_{\mathrm{in}}}
\sum_{\vec a}\sum_{\vec q,\vec q'} A_{\vec q,\vec q'}\,e^{i(\vec q'-\vec q)\cdot\vec a}\nonumber\\
&\qquad \times\sum_{j,j_z}\frac{S^{(\alpha,\beta,\gamma)}_{j,j_z}(\vec q,\vec q',E-\Delta E_{j,j_z})}{q^2 q'^2}.
\label{eq:full_xs}
\end{align}

All atomic matrix elements entering $S_{j,j_z}$ are rotated to the $\vec{N}$-aligned frame using Eq.~\eqref{eq:wigner_rot}.
The factors $A_{\vec{q},\vec{q}'}$ account for the dynamical diffraction amplitudes of elastically scattered electron waves, as defined in Refs.~\cite{Schattschneider2006, Rusz2007First-principlesFerromagnets}.

The rest of the MDFF-based code and its propagation through dynamical diffraction follows the standard formalism described in~\cite{Schattschneider2006, Rusz2007First-principlesFerromagnets}. A detailed implementation of these procedures is provided in the Supp.\ Note~S2~\cite{EMLD-th-suppl}.

\end{document}


\preprint{APS/123-QED}

\title{SUPPLEMENTAL NOTE for: Vectorial Reconstruction of Magnetic Order Parameters using Electron Magnetic Linear Dichroism}

\author{Jan Hajduček}
\email{jan.hajducek@ceitec.vutbr.cz}
\affiliation{CEITEC BUT, Brno University of Technology, Purkyňova 123, 612 00 Brno, Czech Republic}

\author{Jáchym Štindl}%
\affiliation{Institute of Physical Engineering, Brno University of Technology, Technická 2, 616 69 Brno, Czech Republic}

\author{Ján Rusz}%
\affiliation{Department of Physics and Astronomy, Uppsala University, Box 516, 75120 Uppsala, Sweden}

\author{Vojtěch Uhlíř}
\email{vojtech.uhlir@ceitec.vutbr.cz}
\affiliation{CEITEC BUT, Brno University of Technology, Purkyňova 123, 612 00 Brno, Czech Republic}
\affiliation{Institute of Physical Engineering, Brno University of Technology, Technická 2, 616 69 Brno, Czech Republic}


\date{\today}

\maketitle

\section{\label{sec:Results1} Separation and quantification of electron-dichroic components}

Momentum-resolved spectral differences in EELS may contain three distinct contributions of electron dichroism: electron magnetic circular dichroism (EMCD); electron magnetic linear dichroism (EMLD); and the nonmagnetic electron natural linear dichroism (ENLD). Their superposition can be written as:

\begin{align}
    \Delta I (E) &= I_{\Omega_1}(E)-I_{\Omega_2}(E)=\notag\\
    &= \Delta I_\text{EMCD}(E) + \Delta I_\text{EMLD}(E) + \Delta I_\text{ENLD}(E).
\end{align}

\begin{figure*}[t]
\includegraphics[width = 1\columnwidth]{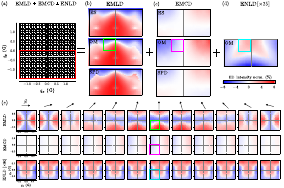}
\caption{\label{figS1:Separation}\textbf{Separation of EMLD, EMCD, and ENLD in simulated FeRh spectra.}(a) L$_3$-edge mean-subtracted spectral profiles over the calculated diffraction plane for one FM Fe sublattice of 10-nm-thick AF FeRh oriented along ZA~$\parallel[016]$ beam geometry with $\vec{M}\parallel[010]$. The profiles contain EMLD, EMCD, and ENLD contributions. (b--d) EMLD, EMCD, and ENLD maps evaluated from the bottom half of the calculated region obtained using the separation procedures summarized in the text. The OM, ES, and SPD methods yield consistent spatial symmetries of EMCD and EMLD, with the largest deviations occurring near the regions of the dichroic signal sign reversal. ENLD is evaluated for the full compensated AF structure using the OM based decomposition with ES off. Relative signal intensity is quantified using Eq.~\ref{eq:NN1} normalized by the averaged spectral intensity $\int_{L_3}\left\langle I_{\vec{q}}(E)\right\rangle_{\vec{q}\in Q}dE$. (e) EMLD, EMCD, and ENLD maps for an in-plane (IP) rotation of the local spin direction in steps of $15^\circ$ evaluated using OM method. EMLD and EMCD redistribute with the magnetic-axis orientation, whereas ENLD remains independent of spin direction. Green, magenta, and cyan squares in panels (b)-(d) indicate corresponding regions in the dichroic maps in panel (e).}
\end{figure*}

Separating these terms is required both for interpreting the simulations and for identifying experimentally suitable collection geometries. We compare three proposed separating procedures: a spectral-profile decomposition (SPD) method, an exchange-splitting (ES) method, and an opposite-magnetization (OM) method. The methods differ in whether they provide a symmetry-exact spectral decomposition, a model-based decomposition, or an approximate scalar separation. Their assumptions and outputs are summarized in Table~\ref{tab:separation-methods} and tested in Fig.~\ref{figS1:Separation}.

At each reciprocal-space position, we define a~mean-subtracted spectral profile:
\begin{equation}
   \delta I_{\vec{q}}(E)=I_{\vec{q}}(E)-\left\langle I_{\vec{q}}(E)\right\rangle_{\vec{q}\in Q},
   \label{eq:deltaI}
\end{equation}
\noindent where $Q$ denotes the calculated reciprocal-space region. This subtraction removes the momentum-averaged spectral component and retains the anisotropic redistribution of intensity across the diffraction plane. The resulting $\delta I_{\vec{q}}(E)$ is the quantity decomposed below. Its numerical baseline depends on the selected region $Q$, so comparisons use the same reciprocal-space region throughout.

For dichroic signal maps in the diffraction plane, we quantify the signed amplitude of a~spectral difference using
\begin{align}
A_{\Delta} &= \chi \int_{L_3} \left|\Delta I(E)\right|\mathrm{d}E,
\label{eq:NN1}
\end{align}
\noindent where $\chi=\pm 1$ is assigned from the initial slope at the onset of the L$_3$ edge and the integration is restricted in the case of Fe L$_3$ edge to 706--711~eV. Equation~\eqref{eq:NN1} defines a~signed signal-amplitude metric rather than a~linear spectral integral: the absolute value prevents cancellation between positive and negative lobes. Consequently, additivity of scalar amplitudes is not generally guaranteed and is invoked below only within the line-shape assumptions of the SPD procedure.

The three procedures are introduced in the order SPD, ES, and OM. The first provides an approximate scalar separation, the second a~model-based decomposition, and the third an odd/even symmetry decomposition.

\vspace{0.3cm}

\textbf{Spectral-profile decomposition (SPD).} The SPD method provides an approximate scalar separation based on the distinct spectral shapes of EMCD and EMLD in metallic $3d$ systems such as FeRh. In the present calculations, EMCD has a~nonzero cumulative integral between the $L_3$ and $L_2$ edges, whereas the oscillatory EMLD profile integrates approximately to zero over the same range:
\begin{align*}
J_C(E_m) &= \int_{L_3}^{E_m} \Delta I_C(E)\, \mathrm{d}E \neq 0, &
J_L(E_m) &= \int_{L_3}^{E_m} \Delta I_L(E)\, \mathrm{d}E \approx 0.
\end{align*}
We therefore estimate the circular and linear scalar amplitudes as
\begin{align}
A_C^{\mathrm{SPD}} &= J_C(E_m), \\
A_L^{\mathrm{SPD}} &= A_{\Delta}-A_C^{\mathrm{SPD}}.
\label{eq:NN2}
\end{align}
\noindent The signs are assigned consistently using the cumulative EMCD integral and the onset-slope convention in Eq.~\eqref{eq:NN1}. This is not a~general spectral decomposition and the relation $A_{\Delta}\approx A_C^{\mathrm{SPD}}+A_L^{\mathrm{SPD}}$ is a~line-shape-based assignment, not a~linear identity. The procedure relies on the EMCD and EMLD profiles remaining sufficiently distinct and should not be applied without validation to systems dominated by multiplets, inequivalent sites, or strong orbital anisotropy.

\vspace{0.3cm}

\textbf{Exchange-splitting (ES) method.} Within the presented spectral calculations, the EMLD contribution is generated by the projection-dependent exchange shifts applied to the $2p$ core sublevels. Repeating the calculation with these shifts set to zero suppresses EMLD while retaining the valence-spin-polarization contribution representing EMCD and the non-magnetic ENLD response. The difference between calculations with and without core-level exchange splitting therefore identifies the EMLD-associated term. Both these components can be formally expressed as:
\begin{align}
\delta I_{\vec q}^{j_{z}~\mathrm{off}}(E) &= \Delta I_{\mathrm{EMCD}} + \Delta I_{\mathrm{ENLD}},\\
\frac{\delta I_{\vec q}^{j_{z}~\mathrm{on}}(E)-\delta I_{\vec q}^{j_{z}~\mathrm{off}}(E)}{2} &= \Delta I_{\mathrm{EMLD}}.
\end{align}

Because core-level exchange splitting also produces smaller changes in the EMCD spectral profile, the ES procedure is a~model-based decomposition rather than a~fully interaction-independent separation.

\vspace{0.3cm}

\textbf{Opposite-magnetization (OM) method.} The assumption that EMCD is odd under reversal of an FM magnetization, $\vec M\rightarrow-\vec M$, whereas EMLD is even, allows using these symmetries for disentangling dichroic components. Formally, we can write:
\begin{align}
\delta I_{\vec q}^{\mathrm{odd}}(E) &= \frac{\delta I_{\vec q}(E;\vec M)-\delta I_{\vec q}(E;-\vec M)}{2}=\Delta I_{\mathrm{EMCD}},\\
\delta I_{\vec q}^{\mathrm{even,on}}(E) &= \frac{\delta I_{\vec q}^{\mathrm{on}}(E;\vec M)+\delta I_{\vec q}^{\mathrm{on}}(E;-\vec M)}{2}=\Delta I_{\mathrm{EMLD}}+\Delta I_{\mathrm{ENLD}}.
\end{align}
Repeating the even combination with the core-level exchange splitting disabled gives the ENLD reference and EMLD follows by subtraction as:
\begin{align}
    \delta I_{\vec q}^{\mathrm{even,off}}(E)&=\Delta I_{\mathrm{ENLD}},\\
    \delta I_{\vec q}^{\mathrm{even,on}}(E) - \delta I_{\vec q}^{\mathrm{even,off}}(E)&=\Delta I_{\mathrm{EMLD}}.
\end{align}

\begin{table}[b]
\centering
\renewcommand{\arraystretch}{1.35}
\caption{\label{tab:separation-methods}Assumptions and applicability of the proposed separation methods for simulated electron dichroism spectra.}
\begin{tabular}{ccccc}
\hline\hline
\textbf{Method} & \parbox[t]{1.35in}{\centering\textbf{Principle}} & \parbox[t]{1.55in}{\centering\textbf{Output}} & \parbox[t]{1.05in}{\centering\textbf{Status}} & \parbox[t]{1.75in}{\centering\textbf{Experimental realization}} \\
\hline
SPD & \parbox[t]{1.35in}{Exploit distinct line shapes} & \parbox[t]{1.55in}{Scalar EMCD/EMLD amplitudes} & \parbox[t]{1.05in}{Approximate} & \parbox[t]{1.55in}{Spectral integration after material-specific validation} \\
ES & \parbox[t]{1.35in}{Switch imposed core exchange shifts on/off} & \parbox[t]{1.55in}{EMLD-associated spectrum within the model} & \parbox[t]{1.05in}{Model-based} & \parbox[t]{1.55in}{Comparing below/above ordering temperature} \\
OM & \parbox[t]{1.35in}{Use odd/even response under $\vec{M}$ reversal} & \parbox[t]{1.55in}{EMCD and EMLD spectra, ENLD with ES off} & \parbox[t]{1.05in}{Symmetry-exact} & \parbox[t]{1.55in}{Magnetization reversal in an FM state} \\
\hline\hline
\end{tabular}
\end{table}

\vspace{0.3cm}

Figure~\ref{figS1:Separation} compares the practical outputs of the three procedures. They reproduce the same dominant spatial symmetries, with the largest differences near sign reversals where the scalar SPD assignment is least stable. The rotating-sublattice dataset provides a~cross-check that the separated EMCD and EMLD maps follow the expected odd and even dependence on the local magnetic-axis orientation.

Figure~\ref{figS1:Separation}(a) shows the mean-subtracted spectral profiles from which the separation is performed. Panels~\ref{figS1:Separation}(b--d) show the comparison of signal maps of EMLD, EMCD and ENLD, respectively obtained by the OM, ES, and SPD procedures. All three methods recover the same dominant spatial symmetries, while the largest discrepancies appear near sign reversals, where the scalar SPD assignment is least stable. The spin axis dependent separated signal in Figure~\ref{figS1:Separation}(e) provides an additional cross-check of the dichroic signal symmetries. As the local spin direction is rotated in steps of $15^\circ$, EMCD and EMLD redistribute accordingly, whereas ENLD remains unchanged.

The separated maps also identify reciprocal-space regions in which a~collection aperture preferentially enhances EMCD or EMLD. Aperture placement therefore optimizes sensitivity but does not by itself constitute a~general decomposition of the three signals. Experimentally, a~nonmagnetic reference may be obtained by suppressing magnetic order, for example by heating above N\'eel temperature $T_\mathrm{N}$ for AF order or Curie temperature $T_\mathrm{C}$ for FM order. Such subtraction is analogous to the ES decomposition but is not strictly identical, because temperature may also change structural anisotropy, spectral broadening, and dynamical diffraction.

The most direct experimental implementation of the symmetry decomposition is available in FM, where spectra acquired for opposite magnetization directions separate the odd EMCD response from the even EMLD + ENLD contribution. EMLD can then be distinguished from ENLD by rotating the magnetic axis at fixed crystal geometry or by comparison with an independently characterized nonmagnetic reference. A larger ENLD contribution is expected in crystals of lower symmetry or stronger orbital anisotropy~\cite{Yamaguchi2018}.

\newpage

\section{\label{Supp:MDFF} Implementation of vectorial EMD calculations}

In the dipole approximation, the mixed dynamic form factor (MDFF) separates into a~symmetric second-rank tensor and an antisymmetric pseudovector. The antisymmetric term is odd under magnetic reversal and produces EMCD. The symmetric tensor contains the isotropic response and an anisotropic part. The anisotropic part may arise from crystal-field or structural anisotropy, producing ENLD, or from the orientation-dependent magnetic response, producing EMLD. Thus, the tensor decomposition separates linear from circular dichroism, whereas an additional magnetic reference or symmetry operation is required to separate EMLD from ENLD.

The implementation assumes collinear magnetic order and neglects valence-band spin--orbit coupling, so the spin-density matrix is diagonal in the local magnetic frame before rotation. Core-level exchange is introduced through effective $j_z$-dependent energy shifts. These assumptions are appropriate for the present FeRh methodological comparison, but they prevent direct applicability to materials with strong valence SOC or noncollinear spin--orbital entanglement.

The double-differential scattering cross section (DDSCS) in the TEM may be written as \cite{Schattschneider2006,Rusz2007First-principlesFerromagnets}

\begin{equation}
\frac{\partial^2\sigma}{\partial\Omega\,\partial E}
=\frac{4\gamma^2}{a_0^2}\frac{k_f}{k_{\mathrm{in}}}\sum_{\vec a}\sum_{\vec q,\vec q'} A_{\vec q,\vec q'} \,e^{i(\vec q'-\vec q)\cdot\vec a}
\; \frac{S(\vec q,\vec q',E)}{q^2 q'^2},
\label{eq:ddcs}
\end{equation}

\noindent where $A_{\vec q,\vec q'}$ encodes the elastic (dynamical) scattering between incoming and outgoing plane-wave components and $S$ is the MDFF defined as:
\begin{equation}
S(\vec q,\vec q',E)=\sum_{i,f}
\langle i|e^{i\vec q\cdot\hat{\mathrm R}}|f\rangle
\langle f|e^{-i\vec q'\cdot\hat{\mathrm R}}|i\rangle
\delta(E_f-E_i-E).
\label{eq:mdff_def}
\end{equation}

\noindent In the dipole approximation assuming $\vec{q}\cdot \hat{\mathrm R} \ll 1$, we can approximate $e^{i\vec q\cdot\hat{\mathrm R}}\approx 1+i\vec q\cdot\hat{\mathrm R}$, with which the MDFF decomposes naturally into a~symmetric (tensor-like) part and an antisymmetric (pseudovector-like) part as \cite{Negi2018ProbingBeams,Rusz2016AberratedAspects}:

\begin{align}
S_{\mathrm{dip}}(\vec q,\vec q',E)&=\vec{q} 
\begin{pmatrix}
\mathcal N_{11} & N_{12}+i{\mathcal M}_{12} & N_{13}+i{\mathcal M}_{13} \\
N_{21}-i{\mathcal M}_{12} & \mathcal N_{22} & N_{23}+i{\mathcal M}_{23} \\
N_{31}-i{\mathcal M}_{13} & N_{32}-i{\mathcal M}_{23} & \mathcal N_{33}
\end{pmatrix}
\vec{q}'=\notag\\
&= \vec q\!\cdot\!\hat{\mathcal N}(E)\!\cdot\!\vec q'
+i\;(\vec q\times\vec q')\!\cdot\!\hat{\mathcal M}(E),
\label{eq:dipole_decomp}
\end{align}

\noindent where $\hat{\mathcal N}(E)$ is the symmetric second-rank tensor and $\hat{\mathcal M}(E)$ denotes the antisymmetric magnetic contribution. The trace of $\hat{\mathcal N}$ gives the isotropic response; its anisotropic part contains both ENLD and EMLD. Equation~\eqref{eq:dipole_decomp} therefore establishes the analogy to XMLD through $\hat{\mathcal N}$ and to XMCD through $\hat{\mathcal M}$, but does not alone separate magnetic from structural linear dichroism.

The MDFF \(S(\vec{q}, \vec{q}^{\,\prime}, E)\) can be expressed in a~multipole expansion of the Coulomb interaction as \cite{Schattschneider2006, Rusz2007First-principlesFerromagnets}

\begin{equation}
\begin{aligned}
S(\vec{q}, \vec{q}^{\,\prime}, E) &= 
\sum_{mm'} \sum_{LMS\,L'M'S'} \sum_{\lambda \mu} \sum_{\lambda' \mu'} 
4\pi i^{\lambda - \lambda'} (2l + 1) \sqrt{[\lambda, \lambda', l', L']} \\
&\quad \times Y_{\mu}^{\lambda} (\vec{q}/q)^{*} Y_{\mu'}^{\lambda'} (\vec{q}^{\,\prime}/q')
\langle j_{\lambda} (q) \rangle_{ELSj} \langle j_{\lambda'} (q') \rangle_{EL'S'j} \\
&\quad \times 
\begin{pmatrix}
l & \lambda & L \\
0 & 0 & 0
\end{pmatrix}
\begin{pmatrix}
l & \lambda' & L' \\
0 & 0 & 0
\end{pmatrix} \\
&\quad \times
\begin{pmatrix}
l & \lambda & L \\
-m & \mu & M
\end{pmatrix}
\begin{pmatrix}
l & \lambda' & L' \\
-m' & \mu' & M'
\end{pmatrix} \\
&\quad \times \sum_{j_z} (-1)^{m + m'} (2j + 1) 
\begin{pmatrix}
l & \frac{1}{2} & j \\
m & S & -j_z
\end{pmatrix}
\begin{pmatrix}
l & \frac{1}{2} & j \\
m' & S' & -j_z
\end{pmatrix} \\
&\quad \times \sum_{\vec{k}n} D^{\vec{k}n}_{LMS} 
\left( D^{\vec{k}n}_{L'M'S'} \right)^{\star} \delta \left( E + E_{nlk} - E_{\vec{k}n} \right)
\end{aligned}
\label{eq:MDFF}
\end{equation}

\noindent where the indices \(l, m\) denote orbital quantum numbers of the initial core state, while \(j, j_z\) are the total angular momentum and its projection. The indices \(L, M, S\) and \(L', M', S'\) label orbital, magnetic, and spin quantum numbers of the final-state basis inside the atomic sphere. The summations over \(m, m'\), and \(j_z\) account for all core-state magnetic sublevels and spin couplings. The spherical harmonics \(Y_\mu^\lambda(\vec{q}/q)\) and \(Y_{\mu'}^{\lambda'}(\vec{q}^{\,\prime}/q')\) describe the angular dependence of the momentum transfer directions, while \(j_\lambda(q)\) and \(j_{\lambda'}(q')\) are spherical Bessel functions arising from the Rayleigh expansion of the plane-wave factor. The quantities \(\langle j_\lambda(q) \rangle_{ELSj}\) denote radial matrix elements coupling core and valence states. The Wigner \(3j\)-symbols impose angular momentum selection rules for orbital coupling (\(l \leftrightarrow L\)) and spin-orbit coupling (\(l \leftrightarrow j\)). The factor \((-1)^{m+m'}(2j+1)\) results from summation over intermediate spin projections~\(j_z\). The coefficients \(D^{\vec{k}n}_{LMS}\) represent projections of Bloch states \(|\vec{k}n\rangle\) onto the local atomic basis \(|LMS\rangle\) inside the atomic sphere. The summation over \(\vec{k}n\) runs over all final Bloch states, while the Dirac delta function ensures energy conservation between the core level energy \(E_{nlk}\) and the final-state eigenenergy \(E_{\vec{k}n}\).

\begin{figure*}[t]
\includegraphics[width = 1\columnwidth]{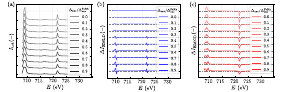}
\caption{\label{figS2:jz}\textbf{Dependence of the calculated dichroic spectra on core-level exchange splitting.} Representative spectra of (a) total EELS intensity, (b) EMLD, and (c) EMCD calculated while scaling the imposed core-level exchange shift $\Delta_{\mathrm{exc}}$ relative to the FeRh value $\Delta_{\mathrm{exc}}^{\mathrm{FeRh}}$.}
\end{figure*}

To include EMLD signal component, core-level exchange splitting is included by energy shifting $j_z$-resolved core states, which is parameterized as $\Delta E_{j,j_z}=\lambda_j j_z$, where $\lambda_j$ is an effective core-level exchange parameter rather than a~quantity obtained self-consistently in the present WIEN2k calculation. The FeRh values are specified in Sec.~\ref{Supp:parameters}, and their influence is tested by scaling the imposed energy shift in Fig.~\ref{figS2:jz}. Figure~\ref{figS2:jz}(a--c) show representative spectra of the total EELS intensity, the EMLD component, and the EMCD component, respectively, as a~function of the imposed core-level exchange shift $\Delta_{\mathrm{exc}}$, scaled relative to the FeRh value $\Delta_{\mathrm{exc}}^{\mathrm{FeRh}}$. As the splitting is increased from zero to the physical FeRh value, the EMLD term grows approximately linearly, whereas the EMCD signal shows smaller amplitude changes and line-shape modifications. Previous work has shown that including exchange splitting in XMCD calculations produces a slight energy shift of the EMCD spectral profile~\cite{Kunes2002X-raySplitting, Ebert1996}. Here we find that exchange splitting alters the EMCD spectral profile, while the diffraction-plane distribution remains unchanged. These results underscore the need for accurate modelling of exchange-split core levels in quantitative interpretations of EMCD and EMLD in FeRh.

Following this formalism, we implement vectorial rotations of both the exchange-splitting axis and the spin-polarization direction using Wigner $D$-matrices in the MDFF calculation code \cite{Rusz2007First-principlesFerromagnets}. 

The exchange-splitting axis is rotated through the Clebsch--Gordan coefficients collected in the $X$-tensor components defined in Eq.~\eqref{eq_Xtensor}, whereas spin polarization is treated by rotating spin-$\tfrac{1}{2}$ states. This provides a~consistent Euler-angle parametrization of the local magnetic-axis orientation. The rotated coupled state is expressed as
\begin{align}
\langle \vec{r}s| j j_z \rangle_{\hat{z} \parallel\vec{M}}
&= \sum_{j_z'} \langle \vec{r}s | j j_z' \rangle \langle j j_z' | j j_z \rangle \notag\\[6pt]
&= \sum_{j_z'} C_{l m \tfrac{1}{2} s}^{j j_z'} \, R_{js}(\vec{r}) \, Y_{lm} (\hat{r})
     \, D^{(j)}_{j_z' j_z}(\alpha,\beta,\gamma).
\end{align}

\noindent where we express Wigner matrices as:

\begin{equation}
    D^{(j)}_{m' m}(\alpha,\beta,\gamma) = e^{-i m' \alpha}\; d^{(j)}_{m' m}(\beta)\; e^{-i m \gamma}.
\end{equation}

\noindent Here, $d^{(j)}_{m' m}$ are so-called reduced Wigner matrices. From this, we compose mixed rotated matrix components $X(\alpha, \beta, \gamma)$ used in the MDFF calculation \cite{Rusz2007First-principlesFerromagnets} as:

\begin{align}
X^{(j j_z)}_{m m'}(\alpha, \beta, \gamma) 
&= \sum_{j_z',j_z''} 
    C_{l m \tfrac{1}{2} s}^{j j_z'} 
    D^{(j)}_{j_z j_z'}(\alpha,\beta,\gamma) 
    C_{l m' \tfrac{1}{2} s}^{j j_z''} 
    D^{(j)*}_{j_z j_z''}(\alpha,\beta,\gamma) =\notag\\[6pt]
&= \sum_{j_z',j_z''} 
    C_{l m \tfrac{1}{2} s}^{j j_z'} 
    C_{l m' \tfrac{1}{2} s}^{j j_z''} \,
    d^{(j)}_{j_z' j_z}(\beta) \, d^{(j)}_{j_z'' j_z}(\beta) \,
    e^{-i (j_z'-j_z'') \alpha}.
    \label{eq_Xtensor}
\end{align}

Here $C_{l m \tfrac{1}{2} s}^{j j_z'}$ represent the Clebsch-Gordon coefficients, the $\beta$-dependence enters through the reduced Wigner functions $d^{(j)}_{m m'}(\beta)$, while the azimuthal angle $\alpha$ appears as a~relative phase between different $m$-channels. As expected, the third Euler angle $\gamma$ cancels out.

The spin-polarization part is treated separately. Under the stated neglect of valence-band spin--orbit coupling, the spin operator is diagonal in the local two-spin subspace $\{|\!\uparrow\uparrow\rangle,|\!\downarrow\downarrow\rangle\}$ before rotation:

\begin{equation}
    M = \begin{pmatrix}
M_{\uparrow\uparrow} & 0 \\[4pt]
0 & M_{\downarrow\downarrow}
\end{pmatrix}.
\end{equation}

Applying the similarity transform with the spin-$\tfrac{1}{2}$ Wigner matrix $D^{(1/2)}(\alpha,\beta,\gamma)$, the rotated two-spin subspace can be expressed as:
\begin{equation}
    M' = D^{(1/2)}(\alpha,\beta,\gamma)\, M \, D^{(1/2)\dagger}(\alpha,\beta,\gamma),
\end{equation}

\noindent which gives:
\begin{equation}
M'(\alpha,\beta) =
\begin{pmatrix}
c^2 M_{\uparrow\uparrow} + s^2 M_{\downarrow\downarrow} &
c s \, e^{-i\alpha}\,(M_{\uparrow\uparrow} - M_{\downarrow\downarrow}) \\[10pt]
c s \, e^{i\alpha}\,(M_{\uparrow\uparrow} - M_{\downarrow\downarrow}) &
s^2 M_{\uparrow\uparrow} + c^2 M_{\downarrow\downarrow}
\end{pmatrix},
\end{equation}

\noindent with $c=\cos(\beta/2)$ and $s=\sin(\beta/2)$. The result is independent of $\gamma$, as required by rotational covariance. For $\alpha=\beta=\gamma=0$ and $\lambda_{1/2}=\lambda_{3/2}=0$, the implementation reduces algebraically to the unrotated formulation of Ref.~\cite{Rusz2007First-principlesFerromagnets}.

Increasing the imposed core-level exchange splitting produces an approximately linear increase in the EMLD amplitude over the investigated range. The EMCD distribution in reciprocal space retains the same dominant symmetry, but its spectral profile and amplitude change weakly because the shifted core sublevels redistribute spectral weight. Related exchange-induced modifications of magnetic dichroic spectra have been discussed for XMCD~\cite{Kunes2002X-raySplitting,Ebert1996}. Thus, switching off the core exchange is effective for identifying the EMLD-associated term within the present model, but it is not completely neutral with respect to the EMCD line shape.

\section{\label{Supp:EMCDorientation} Dependence of EMCD on beam and magnetic-axis orientation}

EMCD depends strongly on both diffraction geometry and the projection of the local magnetic moment. To characterize this dependence independently of AF magnetic moment compensation, we first calculate the EMCD response of one Fe sublattice of AF FeRh, treated as a~locally FM polarized contribution. The corresponding response of the full AF structure cancels when the electron probe resolves the two sublattices equally.

\begin{figure*}[h]
\includegraphics[width = 1\columnwidth]{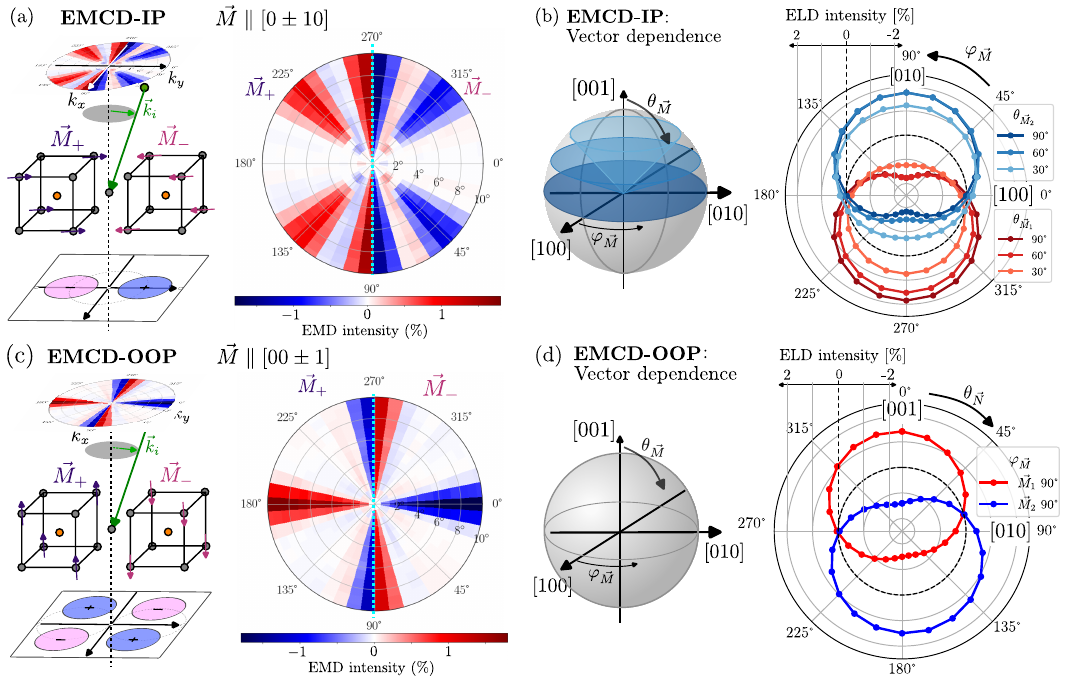}
\caption{\label{figS3:EMCD}\textbf{Dependence of EMCD on beam and magnetic-axis orientation in FeRh.} (a) Geometry for detecting the IP EMCD component from one Fe sublattice of AF FeRh, together with the integrated EMCD signal as a~function of incident-beam tilt magnitude and azimuth relative to the $\text{ZA} \parallel$ [001] zone axis. (b) Dependence of the IP EMCD signal on the azimuthal magnetic angle $\varphi_{\vec M}$ for several polar angles $\theta_{\vec M}$. (c,d) Corresponding results for the OOP-sensitive collection geometry. The finite-signal regions are concentrated near two- and three-beam diffraction conditions.}
\end{figure*}

Figure~\ref{figS3:EMCD}(a) shows the IP-sensitive EMCD signal as a~function of incident-beam direction $\vec{k}_i$ for a~fixed collection geometry. The beam is tilted by up to $10^\circ$ from $\mathrm{ZA}\parallel[001]$ in all azimuthal directions. The signal is negligible over most incident directions and becomes appreciable only for sample crystal tilts representing two- and three-beam conditions. The low-signal sector in the plotted azimuthal range is consistent with previous calculations~\cite{Rusz2011}. Figure~\ref{figS3:EMCD}(b) shows the dependence of integrated EMCD signal on the azimuthal angle of magnetic axis $\varphi_{\vec M}$ for several polar angles $\theta_{\vec M}$. This analysis is enabled by the vectorial formalism presented in Section \ref{Supp:MDFF} and shows proportionality of the integrated EMCD signal to the magnetization projection in the [010] direction.

Figures~\ref{figS3:EMCD}(c,d) present the corresponding out-of-plane (OOP)-sensitive collection geometry. The beam-direction map exhibits fourfold-related signal lobes concentrated near two- and three-beam conditions, and the EMCD amplitude scales with the OOP projection of the local magnetic moment.

Although both IP and OOP components can, in principle, be accessed using suitable geometries, the corresponding high-signal regions occupy a~restricted part of beam-orientation space. EMCD-based vector reconstruction is therefore substantially more sensitive to crystal alignment and diffraction conditions than the EMLD approach considered in the main text.

\newpage

\section{\label{Supp:tensor-comparison} Comparison of energy-resolved and energy-integrated tensor calculations}

Here, ``energy-resolved'' denotes the MDFF calculation with explicit core-level transitions and spectral profiles, proposed in Ref. \cite{Rusz2007First-principlesFerromagnets}, whereas ``energy-integrated tensor'' denotes the effective transition-tensor treatment propagated through dynamical diffraction proposed in Ref. \cite{Rusz2011}. The latter is used to compare signal symmetry and relative thickness trends, not to reproduce an absolute measured spectrum.

In close analogy to XMLD, EMLD originates from anisotropy of the dipole transition-probability tensor. In Eq.~\eqref{eq:dipole_decomp}, $\hat{\mathcal N}(E)$ contains the isotropic response and its symmetric anisotropic part, whereas $\hat{\mathcal M}(E)$ contains the antisymmetric magnetic contribution. The tensor $\hat{\mathcal N}(E)$ is analogous to the complex dielectric tensor $\hat{\epsilon}(E)$ in optical and X-ray absorption.

In the case of XMLD, the anisotropy between the diagonal components of the dielectric tensor $\hat{\epsilon}$ determines the dichroic contrast and can be written as \cite{Kunes2003}:

\begin{equation}
   \text{XMLD} \approx \dfrac{\omega d}{2c\overline{n}}
\text{Im}\!\left[\epsilon_{\parallel}-\epsilon_{\perp}-\dfrac{\epsilon_{\text{od}}^{2}}{\epsilon_{\perp}}\right],
\end{equation}

\noindent where the $\epsilon_{\parallel}$ and $\epsilon_{\perp}$ correspond to axial components of dielectric tensor for X-ray polarizations $\vec{\varepsilon}$ parallel and perpendicular to the magnetization, $\epsilon_{\text{od}}$ is the off-diagonal component of dielectric tensor, $d$ is the thickness of the film and $\omega$ is the frequency of the X-ray radiation.

In electron scattering, the vector $\vec q$ replaces the photon polarization vector $\vec E$, and the symmetric term $\vec q\!\cdot\!\hat{\mathcal N}\!\cdot\!\vec q'$ defines the angular dependence of the non-magnetic transition strength. Thus, the diagonal components of $\hat{\mathcal N}(E)$ act as \emph{effective dielectric constants} for the directions of $\vec q$ in reciprocal space. Any anisotropy in $\mathcal N_{ii}$ with respect to $\mathcal N_{jj}$ or $\mathcal N_{kk}$ gives rise to an~EMLD contrast that is the electron-scattering analogue of XMLD.

For a~qualitative visualization of this anisotropy, following the analogy with XMLD we define the normalized EMLD intensity for a~given principal axis $i$ as:

\begin{equation}
I^{\mathrm{EMLD}}_{i,\mathrm{norm}}(\vec{q})
\approx \dfrac{\mathcal N_{\parallel}(\vec{q})- \mathcal N_{\perp}(\vec{q})}{\mathcal N_{\text{tot}}(\vec{q})} = 
\dfrac{
  \mathcal N_{ii}(\vec{q}) - \tfrac{1}{2}\big[ \mathcal N_{jj}(\vec{q}) + \mathcal N_{kk}(\vec{q})\big]
}{
  \mathcal N_{11}(\vec{q}) + \mathcal N_{22}(\vec{q}) + \mathcal N_{33}(\vec{q})
},
\label{eq:EMLD_norm}
\end{equation}

\noindent where $\mathcal N_{ii}(\vec q)$ denotes the projection of $\hat{\mathcal N}$ along principal axis $i$, and $(i,j,k)$ are cyclic indices. Equation~\eqref{eq:EMLD_norm} defines a~normalized tensor-based contrast measure used to compare reciprocal-space symmetry and relative thickness dependence; it is not an absolute expression for experimentally measured EMLD intensity.

\begin{figure*}[b]
\includegraphics[width = 1\columnwidth]{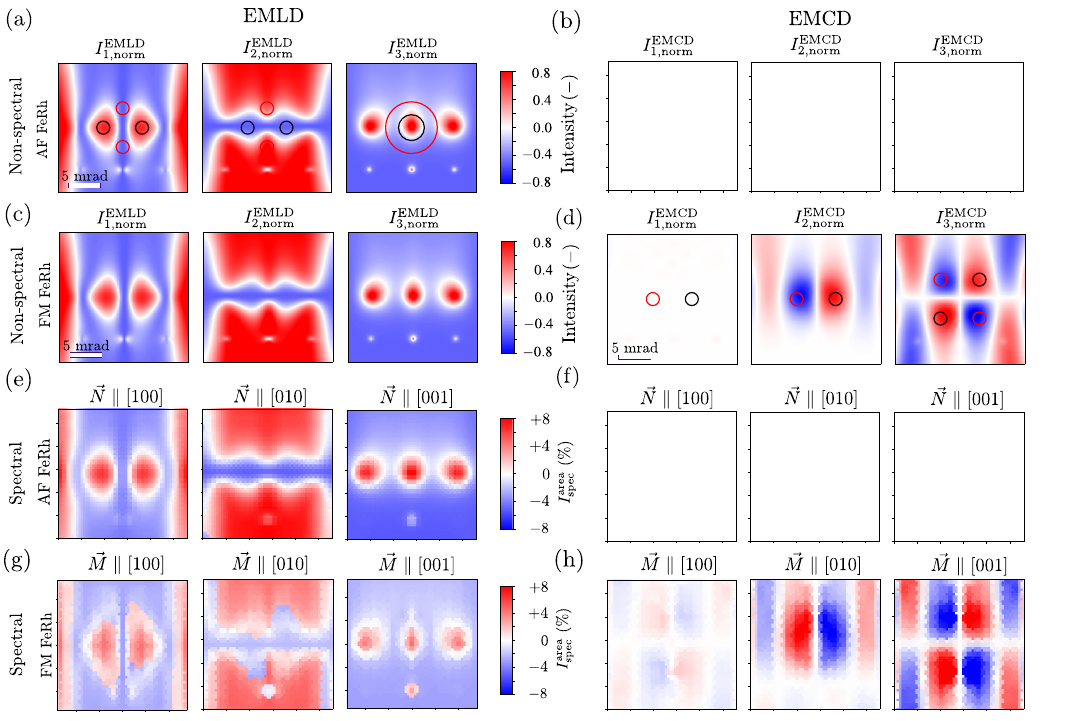}
\caption{\label{figS4:Spec-noSpec}\textbf{Comparison of non-spectral energy-integrated and spectral energy-resolved dichroism calculations.} Reciprocal-space distributions calculated using the energy-integrated tensor approach for (a) EMLD and (b) EMCD in the AF phase, and for (c) EMLD and (d) EMCD in the FM phase of FeRh, respectively. Each panel represents dichroic maps for the Cartesian magnetic-axis components along $x$, $y$, and $z$, respectively. (e,f) Corresponding signal maps obtained from energy-resolved spectral simulations for AF FeRh, yielding great agreement in EMLD and expected net zero EMCD. (g,h) Energy-resolved results for FM FeRh reveal highly similar EMLD and EMCD signal distribution as obtained using non-spectral approach. Circles in (a) and (d) indicate the integration regions used for the thickness dependence of dichroic contrasts in Fig.~\ref{figS5:Thc-dep-noSpec}. All maps are evaluated for FeRh with the thickness of $t_{\mathrm{FeRh}}=30$~nm.}
\end{figure*}

After separating the isotropic trace from the symmetric transition tensor, the remaining rank-2 anisotropic component is traceless. Consequently, the three cyclic EMLD contrasts defined in Eq.~\eqref{eq:EMLD_norm} are not independent and satisfy

\begin{equation}
I^{\mathrm{EMLD}}_{1,\mathrm{norm}} +
I^{\mathrm{EMLD}}_{2,\mathrm{norm}} +
 I^{\mathrm{EMLD}}_{3,\mathrm{norm}}=0.
\end{equation}

\noindent This mirrors the well-known constraint in XMLD: the dichroic signal redistributes intensity among principal axes without introducing any isotropic contribution. Consequently, only two independent EMLD contrast functions exist, consistent with the traceless nature of the underlying anisotropic response \cite{Brouder1990AngularSpectra, Kunes2003}.

In the case of EMCD, the dichroic contrast originates from the antisymmetric part of the MDFF, which is proportional to the imaginary components of the transition tensor.
Analogously to the EMLD definition above, we introduce a~normalized EMCD intensity that isolates the magnetic contribution associated with the off-diagonal tensor components:

\begin{equation}
I^{\mathrm{EMCD}}_{1,2,3,\mathrm{norm}}(\vec{q})
\propto
\dfrac{
\mathcal M_{23,13,12}(\vec{q})
}{
\mathcal N_{11}(\vec{q}) + \mathcal N_{22}(\vec{q}) + \mathcal N_{33}(\vec{q})
},
\label{eq:EMCD_norm}
\end{equation}

\noindent where $\mathcal M_{23}$, $\mathcal M_{13}$, and $\mathcal M_{12}$ denote the imaginary cross terms associated with the three Cartesian moment components. As for Eq.~\eqref{eq:EMLD_norm}, Eq.~\eqref{eq:EMCD_norm} is a~normalized tensor-based contrast measure for comparing symmetry and relative trends, not an absolute measured intensity. Together, Eqs.~\eqref{eq:EMLD_norm} and~\eqref{eq:EMCD_norm} highlight the complementary symmetric and antisymmetric contributions to linear and circular dichroism.

Figure~\ref{figS4:Spec-noSpec} compares the energy-integrated tensor and energy-resolved calculations obtained for both magnetic phases of FeRh. The dichroic signal maps obtained from dielectric tensor formalism in panels (a,b) and (c,d) are a good representation of the AF phase and FM phase, respectively. The signal maps from energy-resolved calculations are shown for AF FeRh in Fig.~\ref{figS4:Spec-noSpec}(e,f) and for FM FeRh in Fig.~\ref{figS4:Spec-noSpec}(g,h).

\begin{figure*}[h]
\includegraphics[width = 1\columnwidth]{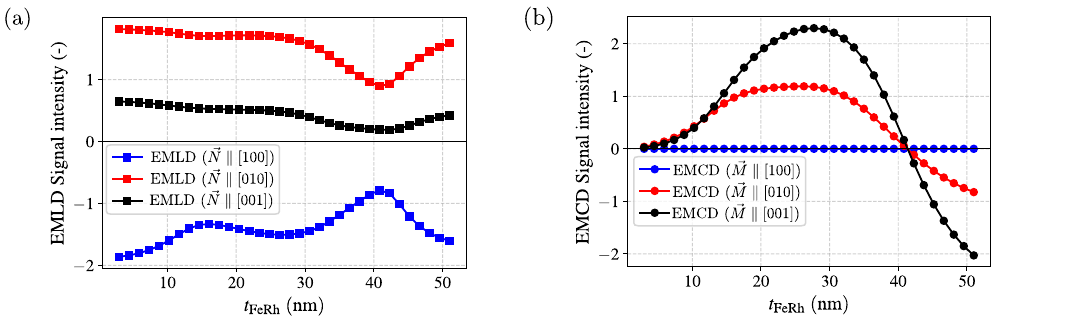}
\caption{\label{figS5:Thc-dep-noSpec}\textbf{Thickness dependence of EMLD and EMCD.} Normalized amplitudes of the $x$, $y$, and $z$ components of (a) EMLD in AF FeRh and (b) EMCD in FM FeRh, respectively, as a function of film thickness. The signals were calculated using the energy-integrated tensor approach and the collection regions marked in Fig.~\ref{figS4:Spec-noSpec}. EMLD varies comparatively smoothly with thickness, whereas EMCD shows oscillation and sign reversal.}
\end{figure*}

For the EMLD signal, the two approaches yield the same nodal structure and lobe orientation, with closely matching reciprocal-space distributions. Similarly good agreement between spectral and non-spectral approaches is obtained for EMCD in the FM phase. Overall, we can state that the dominant dichroic signal portion is already well represented by non-spectral calculations for FeRh; however, the spectral quantification and fully vectorial formalism remains limited to the full ab-initio spectral approach. The circular apertures in Fig.~\ref{figS4:Spec-noSpec}(a) and (c) mark the reciprocal-space regions used to extract the thickness dependencies shown in Fig.~\ref{figS5:Thc-dep-noSpec} of EMLD in the AF phase and EMCD in the FM phase, respectively.

Thickness-dependent dichroic signals in Figure~\ref{figS5:Thc-dep-noSpec} show that EMCD undergoes oscillation and sign reversal with thickness because its antisymmetric contribution depends sensitively on phase interference between elastic-scattering pathways. EMLD is also modified by dynamical diffraction, as evidenced by the dip near 42~nm, but its overall amplitude and sign are substantially more stable over the investigated thickness range. Overall, EMLD is less sensitive to thickness-dependent interference effects.

The comparison supports use of the energy-integrated tensor model for signal symmetry and relative thickness trends, whereas the energy-resolved MDFF calculation remains necessary for spectral profiles and quantitative sublattice-resolved interpretation.

\newpage
\section{\label{Supp:probe-confinement} Probe-confinement effect on electron dichroism}

The vectorial sensitivity of EMLD becomes particularly useful with spatially confined electron probes. For weak beam convergence ($\alpha<2$~mrad), parallel-beam models provide a~good description~\cite{Thersleff2015QuantitativeDichroism}; with increasing convergence, finite-probe effects and dynamical diffraction become more important~\cite{Loffler2018Convergent-beamApplications,Rusz2017LocalizationResolution}.

To examine a~confinement range inaccessible to the parallel-beam spectral calculations, we use the established convergent-beam dynamical-diffraction formalism for EMCD~\cite{Rusz2013,Rusz2017} and introduce EMLD through the corresponding symmetric transition-tensor representation from Eq.~\eqref{eq:EMLD_norm}. This calculation tests persistence and spatial scaling of the linear-dichroic contrast rather than its detailed energy-resolved spectral profile.

\begin{figure*}[h]
\includegraphics[width = 1\columnwidth]{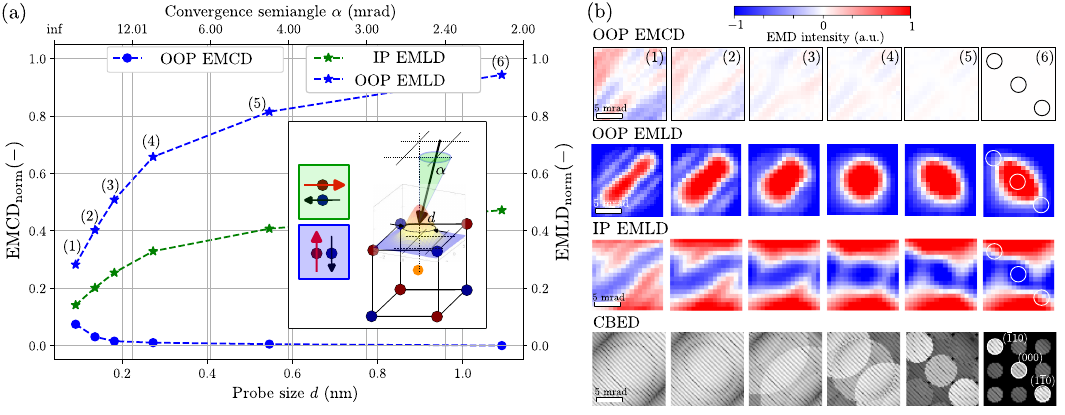}
\caption{\label{figS6:conv}\textbf{Dependence of electron dichroism on probe confinement.} (a) IP and OOP EMLD amplitudes and OOP EMCD amplitude for the full AF lattice as functions of the convergence semiangle and the corresponding aberration-free probe diameter for 10-nm-thick FeRh at 300~keV, with the incident direction near $[1\,1\,13]$. (b) Representative momentum-resolved dichroic maps for probe sizes (1)-(6) indicated in (a), ranging from 0.09 nm to 1.09 nm, approaching the parallel beam condition. The corresponding CBED patterns are shown in the bottom row. Empty circles in the dichroic maps indicate the positions of characteristic Bragg peaks denoted in the CBED pattern. Increasing beam convergence causes progressive overlap of the CBED disks, yielding a modulated dichroic response, yet the EMLD response remains finite down to atomically sized probes.}
\end{figure*}


Figure~\ref{figS6:conv}(a) summarizes the calculated dependence on convergence semiangle for a~10-nm-thick FeRh crystal at 300~keV, with the incident direction near $[1\,1\,13]$. Aberrations are neglected, and the bottom axis reports the corresponding aberration-free probe Rayleigh diameter used in the simulation. Figure~\ref{figS6:conv}(b) shows the momentum-resolved signals and CBED patterns as the diffraction disks progressively overlap.

EMLD remains finite over the full simulated convergence range and decreases by approximately a~factor of two at $\alpha=12$~mrad. This shows that confinement to near-atomic probe dimensions does not intrinsically suppress linear-dichroic sensitivity to AF order. By contrast, EMCD of the compensated AF becomes appreciable only when the probe is sufficiently confined to resolve the two sublattices unequally. EMLD is therefore particularly scalable across a~broad range of probe dimensions.

\section{\label{Supp:parameters} Electronic-structure inputs and electron-scattering parameters}

The electronic structure of the AF and FM phases of FeRh was calculated using WIEN2k~\cite{Blaha2020WIEN2k:Solids}, with parameters summarized in Table~\ref{tab:DFT-par}. AF order was initialized by assigning opposite spin polarizations to the two Fe sublattices. Valence-band spin--orbit coupling was omitted to retain a~diagonal spin basis and isolate the role of core-level exchange splitting in the EMLD response.

\begin{table}[h]
\centering
\caption{\label{tab:DFT-par}WIEN2k parameters used for the AF and FM electronic-structure calculations. The Bohr radius is denoted by $a_0$.}
\begin{ruledtabular}
\begin{tabular}{lcc}
\textbf{Parameter} & \textbf{FeRh (AF)} & \textbf{FeRh (FM)} \\
\hline
Crystal prototype & B2 (CsCl) & B2 (CsCl) \\
Space group & $Pm\bar{3}m$ & $Pm\bar{3}m$ \\
Simulation cell & $2\times2\times2$ AF supercell & $1\times1\times1$ primitive cell \\
Lattice constant $a$ (\AA) & 2.989 & 2.989 \\
$R_{\mathrm{MT}}$(Fe) ($a_0$) & 2.43 & 2.43 \\
$R_{\mathrm{MT}}$(Rh) ($a_0$) & 2.43 & 2.43 \\
$R_{\mathrm{MT}}K_{\max}$ & 8.50 & 8.50 \\
Total $k$-points (per BZ) & 10000 & 1000 \\
Basis functions/atom & $>75$ & $>75$ \\
XC functional & PBE-GGA & PBE-GGA \\
Charge-density convergence (e$^{-}$/unit cell) & $10^{-5}$ & $10^{-5}$ \\
\end{tabular}
\end{ruledtabular}
\end{table}

\begin{figure*}[b]
\includegraphics[width = 1\columnwidth]{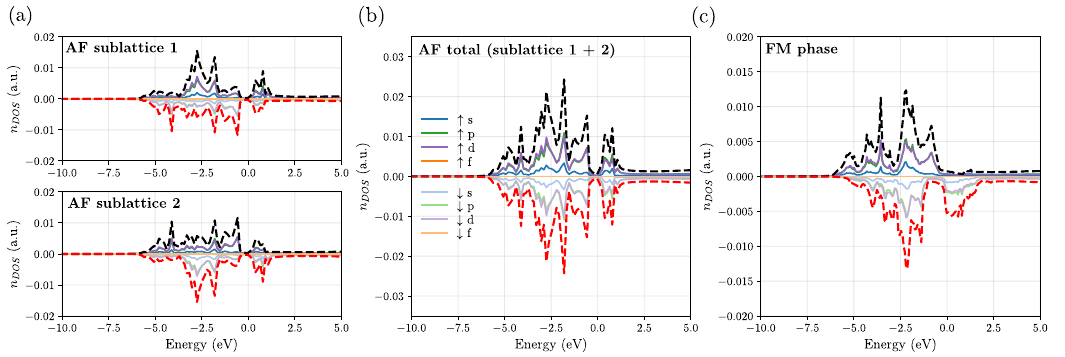}
\caption{\label{figS7:AF-FM-DOS}\textbf{Spin- and orbital-resolved Fe density of states used in the scattering calculations.} (a) DOS of the two oppositely polarized Fe sublattices in AF FeRh calculated without valence-band spin--orbit coupling. (b) Sublattice-summed AF DOS, showing cancellation of the net spin polarization. (c) DOS of FM FeRh, showing a~finite spin imbalance. Positive and negative ordinates denote spin-up and spin-down contributions. Dashed lines show the total DOS in each spin channel, and energy is referenced to the Fermi level.}
\end{figure*}

Figure~\ref{figS7:AF-FM-DOS} shows the orbital- and spin-resolved Fe DOS used as input to the scattering calculations. The two AF Fe sublattices exhibit opposite spin polarizations and cancel upon summation, whereas the FM phase retains a~finite spin imbalance. In both phases, Fe $3d$ states dominate near the Fermi level. These inputs therefore provide compensated and uncompensated spin populations for the AF and FM MDFF calculations, respectively.

For the figures in the main text and Figs.~\ref{figS1:Separation}-\ref{figS3:EMCD}, the energy-resolved MDFF calculations used the 200~keV parameter set stated in the main-text simulation note, a~$2\times2\times2$ AF supercell, a~$0.2G$ reciprocal-space step, an energy step of 0.136~eV, and 0.2~eV Lorentzian broadening. The imposed exchange parameters were $\lambda_{j=3/2}=0.3176$~eV and $\lambda_{j=1/2}=-0.3176$~eV.

For Figs.~\ref{figS4:Spec-noSpec} and~\ref{figS5:Thc-dep-noSpec}, the energy-resolved and energy-integrated calculations were done at 300~keV. The tensor calculation employed the MATS code~\cite{Rusz2013} with the MATS.v2 summation scheme~\cite{Rusz2017}, a reciprocal-space range of $\pm10$~mrad in $q_x$ and $q_y$, 0.25~mrad sampling, a 1.49~nm thickness step, and a Bloch-wave excitation threshold of $10^{-4}$. The crystal was oriented in a three-beam geometry with $\vec G=\pm(100)$ and incident beam near $[016]$.

\bibliography{references-suppl-MAN}